\documentclass[conference,romanappendices]{IEEEtran}
\IEEEoverridecommandlockouts

\usepackage{algorithm,algorithmic,amsmath,amssymb,amsthm,bbm,cite,color,graphicx,microtype,url}
\usepackage[USenglish]{babel}
\usepackage[utf8]{inputenc}
\usepackage[T1]{fontenc}
\usepackage{hyperref}

\usepackage{tikz,pgfplots}
\usetikzlibrary{arrows, arrows.meta, backgrounds, calc, intersections, decorations.markings, decorations.pathreplacing, positioning, shapes}
\pgfplotsset{compat=1.17}

\usepackage{caption}
\renewcommand{\k}{\mathbf{k}}

\newcommand{\Real}{\mbox{$\mathbb{R}$}}
\newcommand{\Compl}{\mbox{$\mathbb{C}$}}

\newcommand{\Diag}{\mathrm{Diag}}

\renewcommand{\Im}{\mathrm{Im}}

\renewcommand{\Pr}{\mathbb{P}}

\renewcommand{\Re}{\mathrm{Re}}
\newcommand{\sgn}{\mathrm{sgn}}

\newcommand{\tran}{\mathrm{T}}

\newtheorem{theorem}{Theorem}

\newtheorem{remark}{Remark}

\newcommand{\Trans}{{\mathrm{T}}}
\newcommand{\Herm}{{\mathrm{H}}}
\newcommand{\trace}{{\mathrm{tr}}}
\renewcommand{\sgn}{\mathrm{sgn}}
\newcommand{\VEC}{\mathrm{vec}}
\renewcommand{\Diag}[1]{\mathtt{diag}{{#1}}}
\newcommand{\BDiag}[1]{\mathtt{DIAG}{{#1}}}

\usepackage{setspace}
\usepackage{fancyhdr}
\fancypagestyle{firstpage}{
  
  \fancyhead[R]{{\scriptsize 1}}
  \fancyfoot[C]{{\footnotesize \begin{singlespace} \textcopyright 2024 IEEE. Personal use of this material is permitted. Permission from IEEE must be obtained for all other uses, in any current or future media, including reprinting/republishing this material for advertising or promotional purposes, creating new collective works, for resale or redistribution to servers or lists, or reuse of any copyrighted component of this work in other works. \end{singlespace}}}}

\title{Optimality of the Bussgang Linear MMSE Channel Estimator for MIMO Systems with 1-Bit ADCs}
\author{
\IEEEauthorblockN{Minhua Ding,$^{1}$ Italo~Atzeni,$^{1}$ Antti~Tölli,$^{1}$ and A.~Lee~Swindlehurst$^{2}$} \\ \vspace{-3mm}
\IEEEauthorblockA{$^{1}$Centre for Wireless Communications, University of Oulu, Finland \\
$^{2}$Department of Electrical Engineering and Computer Science, University of California, Irvine, USA \\
E-mail: \{minhua.ding, italo.atzeni, antti.tolli\}@oulu.fi, swindle@uci.edu}
\thanks{This work was supported by the Research Council of Finland (336449 Profi6, 346208 6G~Flagship, 348396 HIGH-6G, and 357504 EETCAMD), by the European Commission (101095759 Hexa-X-II), and by the U.~S.~National Science Foundation (CCF-2225575).} \vspace{-1mm}}

\begin{document}

\maketitle

\thispagestyle{firstpage}

\begin{abstract}
In this paper, we study the optimality of the Bussgang linear minimum mean squared error (BLMMSE) channel estimator for multiple-input multiple-output systems with 1-bit analog-to-digital converters. We compare the BLMMSE with the optimal minimum mean squared error (MMSE) channel estimator, which is generally non-linear, and we develop a novel framework based on the orthant probability of a multivariate normal distribution to compute the MMSE channel estimate. Then, we analyze the equivalence of the MMSE and BLMMSE channel estimators under specific assumptions on the channel correlation or pilot symbols. Interestingly, the BLMMSE channel estimator turns out to be optimal in several specific cases. Our study culminates with the presentation of a necessary and sufficient condition for the BLMMSE channel estimator to be optimal.
\end{abstract}


\vspace{-1mm}

\section{Introduction} \label{sec:Intro}

Massive multiple-input multiple-output (MIMO) technology is a key enabler of current fifth-generation cellular networks and is envisioned to assume an even more prominent role in future wireless systems~\cite{Raj20}. The large number of antennas in massive MIMO motivates the study of the impact of low-complexity, energy-efficient hardware on the system performance. Low-resolution and 1-bit analog-to-digital converters (ADCs) are considered a promising solution to decrease the complexity and power consumption of fully digital massive MIMO systems~\cite{Jacobsson_TWC_UL, Y_Li_et_al_BLMMSE, Atzeni_2022, Atz21b}. Since coherent MIMO communications require instantaneous channel state information for the beamforming design, the subject of channel estimation with 1-bit ADCs has received special attention in recent years~\cite{Y_Li_et_al_BLMMSE, Atzeni_2022, Utschick_etal_2023}. As a chief example, the Bussgang linear minimum mean squared error (BLMMSE) channel estimator proposed in~\cite{Y_Li_et_al_BLMMSE} uses the second-order statistics~\cite{Bussgang_paper} of 1-bit quantized signals and its linear structure has inspired subsequent signal processing methods applied to, e.g., data detection and beamforming~design.

In general, the optimal minimum mean squared error (MMSE) channel estimator is a non-linear function of the 1-bit quantized signal received during the pilot transmission. Nonetheless, under certain assumptions, the MMSE channel estimator may assume a linear structure and thus be equivalent to its BLMMSE counterpart; in this setting, the BLMMSE channel estimator is clearly optimal in terms of mean squared error (MSE). Recently, it was revealed in~\cite{Utschick_etal_2023} that the MMSE and BLMMSE channel estimators are equivalent in single-input single-output systems with a single pilot symbol or with specifically chosen pilot symbols in the noiseless case. Given the widespread use of the BLMMSE channel estimator, it is of great importance to analyze its optimality in general scenarios, and this is precisely the goal of this work.

In this paper, we develop a new framework based on the orthant probability of a multivariate normal (MVN) distribution to compute the MMSE channel estimate for MIMO systems with 1-bit ADCs. Building on this, we identify specific~cases where the BLMMSE channel estimator is optimal. Then, we present a closed-form expression for the non-linear MMSE channel estimator in a single-input multiple-output (SIMO) system with three receive antennas. Although far from the massive MIMO regime that motivates the use of 1-bit ADCs, studying these specific cases paves the way for establishing a necessary and sufficient condition for the optimality of the BLMMSE channel estimator.\footnote{A more detailed analysis, complemented by additional specific cases and full derivations, can be found in the longer version of this paper~\cite{Din24}.}

\textit{Notation:} $\otimes$ denotes the Kronecker product; $| \cdot |$ represents the determinant of a square matrix or the absolute value of a scalar; $\VEC(\cdot)$ denotes vectorization; $\trace(\cdot)$ represents the trace; $\sgn (\cdot)$ denotes the sign function; $\Re(\cdot)$ and $\Im(\cdot)$ represent the real and imaginary parts, respectively; $j=\sqrt{-1}$ is the imaginary unit; $\Diag(\mathbf{z})$ is a diagonal matrix with the elements of $\mathbf{z}$ on its diagonal; $\widetilde{\Diag{}}(\mathbf{Z})$ is a diagonal matrix with the diagonal elements of the square matrix $\mathbf{Z}$ on its diagonal; $\BDiag(\mathbf{Z}_1, \ldots, \mathbf{Z}_n)$ is a block-diagonal matrix with blocks $\mathbf{Z}_1, \ldots, \mathbf{Z}_n$.

\section{System Model and Problem Statement} \label{sec: sys_model_prob_state}

Consider a general MIMO system with $N_T$ transmit antennas and $N_R$ receive antennas, where $\mathbf{H}\in\mathbb{C}^{N_R \times N_T}$ denotes the channel matrix between the transmitter and receiver. For the channel estimation, $\tau$ pilot symbols per transmit antenna are transmitted, which are collectively denoted by $\mathbf{S}^\Trans = [\mathbf{s}_1 \; \ldots \; \mathbf{s}_\tau] \in \mathbb{C}^{N_T \times \tau}$. The received signal at the input of the ADCs is given by
\begin{align}
\mathbf{B} = \mathbf{H} \mathbf{S}^\Trans + \mathbf{N}, \label{sm: mat_SM}
\end{align}
where $\mathbf{N} \in \mathbb{C}^{N_R \times \tau}$ represents noise. Furthermore, we introduce
\begin{align}
\mathbf{b} = \VEC(\mathbf{B}) = \mathbf{A} \mathbf{h} + \mathbf{n} \in \mathbb{C}^{\tau N_R}, \label{sm: vec_SM}
\end{align}
with
\begin{align}
\mathbf{A} = \mathbf{S} \otimes \mathbf{I}_{N_R} \in \mathbb{C}^{(\tau N_R) \times (N_TN_R)}, \label{eqn: matrix_A}
\end{align}
$\mathbf{h} = \VEC(\mathbf{H}) \in \mathbb{C}^{N_TN_R}$, and $\mathbf{n} = \VEC(\mathbf{N}) \in \mathbb{C}^{\tau N_R}$. We assume a general correlated Rayleigh fading channel model with $\mathbf{h} \sim \mathcal{CN}(\mathbf{0}, \boldsymbol{\Sigma})$, where $\boldsymbol{\Sigma} \in \Compl^{(N_TN_R) \times (N_TN_R)}$ is the channel covariance matrix. The probability density function for $\mathbf{h}$ is expressed as
\begin{align}
p(\mathbf{h})=\frac{1}{\pi^{N_T N_R}|\boldsymbol{\Sigma}|} e^{-\mathbf{h}^\Herm\boldsymbol{\Sigma}^{-1}\mathbf{h}}. \label{eqn: prob_density_h}
\end{align}
Moreover, we assume $\mathbf{n}\sim\mathcal{CN}(\mathbf{0}, \sigma^2\mathbf{I}_{\tau N_R})$, and the resulting signal-to-noise ratio (SNR) is $\textrm{SNR}=\trace(\mathbf{S\mathbf{S}^\Herm})/(\tau N_T\sigma^2)$, which reduces to $|s|^2/\sigma^2$ when $\tau=N_T=1$.

We denote the element-wise memoryless 1-bit quantization operation as
\begin{align}
\mathcal{Q}_{\textrm{1-bit}}(\cdot) = \sgn\big(\Re(\cdot)\big)+j\sgn\big(\Im(\cdot)\big).
\label{eqn: sm_QUANT}
\end{align}
The quantized received signal (at the output of the ADCs) is given by
\begin{align}
\mathbf{r} =\mathcal{Q}_{\textrm{1-bit}}(\mathbf{b})\in\mathbb{C}^{\tau N_R}.
\label{eqn: quantized_y}
\end{align} 
Based on the quantized received signal $\mathbf{r}$, the MMSE estimate of $\mathbf{h}$ is given by $\hat{\mathbf{h}}_{\rm MMSE} = \mathbb{E}(\mathbf{h} | \mathbf{r}) = \int_{\mathbb{C}^{N_T N_R}} \mathbf{h} p(\mathbf{h} | \mathbf{r}) {\rm d} \mathbf{h}$, with $p(\mathbf{h}|\mathbf{r})=\Pr(\mathbf{r}|\mathbf{h})p(\mathbf{h})/\Pr(\mathbf{r})$. Consequently, we have
\begin{align}
\hat{\mathbf{h}}_{\rm MMSE} = \frac{\displaystyle \int_{\mathbb{C}^{N_T N_R}} \mathbf{h} \Pr(\mathbf{r} | \mathbf{h}) p(\mathbf{h}) {\rm d} \mathbf{h}}{\Pr(\mathbf{r})}, \label{eqn: h_MMSE_txt_formula}
\end{align}
with \begin{align}
\Pr(\mathbf{r})= \int_{\mathbb{C}^{N_T N_R}} \Pr(\mathbf{r} | \mathbf{h}) p(\mathbf{h}) {\rm d} \mathbf{h}. \label{eqn: prob_r_first_appear}
\end{align}
An explicit expression for the solution to \eqref{eqn: h_MMSE_txt_formula} is in general unavailable. In the next section, we first develop a novel framework for computing \eqref{eqn: h_MMSE_txt_formula}. Then, in Section~\ref{sec: cases_optimality}, we study some specific cases and determine a general condition under which the MMSE channel estimate is linear in $\mathbf{r}$, i.e., when the BLMMSE channel estimator is optimal.

\section{A Framework for 1-Bit MMSE Channel Estimation} \label{sec: framework}

In this section, we analyze the structure of the MMSE channel estimator with 1-bit ADCs and reformulate the corresponding conditional mean expression into a form that involves the orthant probability of a MVN distribution. This will then be exploited to determine the optimality of the BLMMSE channel estimator.

Let us recall \eqref{sm: vec_SM} and \eqref{eqn: sm_QUANT}--\eqref{eqn: quantized_y}, and let $\mathbf{r}=\mathbf{r}_{\mathsf{R}}+j\mathbf{r}_{\mathsf{I}}$, with 
\begin{align}
\mathbf{r}_{\mathsf{R}}=\Re(\mathbf{r})=\sgn\left(\Re\left(\mathbf{b}\right)\right), \quad \mathbf{r}_{\mathsf{I}}=\Im(\mathbf{r})=\sgn\left(\Im\left(\mathbf{b}\right)\right). \label{eqn:quantized_output_vector}
\end{align}
Then, we write the $k$th element of $\mathbf{r}_{\mathsf{R}}$ and $\mathbf{r}_{\mathsf{I}}$ as $r_{\mathsf{R}, k}=\Re(r_k)$ and $r_{\mathsf{I}, k} = \Im(r_k)$, respectively, with $r_{\mathsf{R}, k}, r_{\mathsf{I}, k} \in \{\pm 1\}$. Correspondingly, we write the $k$th element of $\mathbf{r}$ as
\begin{align}
\hspace{-1mm} r_k & = r_{\mathsf{R}, k} + j r_{\mathsf{I}, k} = \sgn(\Re(b_k)) + j \sgn(\Im(b_k)) \\
& = \sgn(\Re(\mathbf{a}_k^\Trans \mathbf{h}) + \Re(n_k)) + j \sgn(\Im(\mathbf{a}_k^\Trans \mathbf{h}) + \Im(n_k)),
\end{align} 
where $\mathbf{a}_k^\Trans$ denotes the $k$th row of $\mathbf{A}$, $\Re(n_k)$ and $\Im(n_k)$ are independent and identically distributed (i.i.d.) $\mathcal{N}(0, \frac{\sigma^2}{2})$ random variables, and
\begin{align}
(\Re(\mathbf{a}_k^\Trans \mathbf{h}))^2 + (\Im(\mathbf{a}_k^\Trans \mathbf{h}))^2 = \mathbf{h}^\Herm \mathbf{a}_k \mathbf{a}_k^\Trans \mathbf{h}.
\end{align}
It can be readily shown that
\begin{align}
\hspace{-2mm}\Pr(r_{\mathsf{R}, k}|\mathbf{h}) & = \int\limits_0^\infty \frac{\exp\left\{-\frac{1}{\sigma^2}\left(x_k-r_{\mathsf{R}, k}\Re(\mathbf{a}_k^\Trans\mathbf{h})\right)^2\right\}}{\sqrt{\pi\sigma^2}}{\rm d}x_k, \\
\Pr(r_{\mathsf{I}, k}|\mathbf{h}) & = \int\limits_0^\infty \frac{\exp\left\{-\frac{1}{\sigma^2}\left(y_k-r_{\mathsf{I}, k}\Im(\mathbf{a}_k^\Trans\mathbf{h})\right)^2\right\}}{\sqrt{\pi\sigma^2}}{\rm d}y_k.
\end{align}

Based on the distribution of the noise, we have $\Pr(r_{k}|\mathbf{h})=\Pr(r_{\mathsf{R}, k}|\mathbf{h})\Pr(r_{\mathsf{I}, k}|\mathbf{h})$ and, hence,
\begin{align}
&\Pr(\mathbf{r}|\mathbf{h}) = \prod_{k=1}^{\tau N_R}\Pr(r_{k}|\mathbf{h}). \label{eqn: prob_r_given_h_1}
\end{align}
Define $\mathbf{x} = [x_1 \; \ldots \; x_{\tau N_R}]^\Trans$, $\mathbf{y} = [y_1 \; \ldots \; y_{\tau N_R}]^\Trans$, and
\begin{align}
\boldsymbol{\Lambda}_{\mathsf{R}} = \Diag(\mathbf{r}_{\mathsf{R}}), \quad \boldsymbol{\Lambda}_{\mathsf{I}} = \Diag(\mathbf{r}_{\mathsf{I}}). \label{eqn:quantized_output_diag_mat}
\end{align}
Clearly, we have $\boldsymbol{\Lambda}_{\mathsf{R}} \boldsymbol{\Lambda}_{\mathsf{R}} = \boldsymbol{\Lambda}_{\mathsf{I}} \boldsymbol{\Lambda}_{\mathsf{I}} = \mathbf{I}_{\tau N_R}$. Furthermore, it can be shown that \eqref{eqn: prob_r_given_h_1} is equivalent to
\begin{align}
\hspace{-2mm} \Pr(\mathbf{r}|\mathbf{h})= \int_{\mathbb{R}_+^{\tau N_R}}\int_{\mathbb{R}_+^{\tau N_R}} \! \! \frac{\exp\left\{-\frac{\|\boldsymbol{\Lambda}_{\mathsf{R}}\mathbf{x}+j\boldsymbol{\Lambda}_{\mathsf{I}}\mathbf{y} - \mathbf{A}\mathbf{h}\|^2}{\sigma^2}\right\}}{(\pi\sigma^2)^{\tau N_R}}{\rm d}\mathbf{x}{\rm d}\mathbf{y}. \label{eqn: Prob_r_given_h_compact}
\end{align}

Now, introduce the covariance matrix of $\mathbf{b}$ in \eqref{sm: vec_SM}, i.e.,
\begin{align}
\boldsymbol{\Omega}_{\rm b}= \mathbf{A}\boldsymbol{\Sigma}\mathbf{A}^\Herm+\sigma^2\mathbf{I}_{\tau N_R}\in\mathbb{C}^{(\tau N_R)\times(\tau N_R)}, \label{eqn: corr_mat_of_b}
\end{align}
and define
\begin{align}
\mathbf{D}_{\mathsf{R}} & = \Re\left(\boldsymbol{\Omega}_{\rm b}^{-1}\right) = \mathbf{D}_{\mathsf{R}}^\Trans \in\mathbb{R}^{(\tau N_R)\times(\tau N_R)}, \label{eq:D_R} \\
\mathbf{D}_{\mathsf{I}} & = \Im\left(\boldsymbol{\Omega}_{\rm b}^{-1}\right) = -\mathbf{D}_{\mathsf{I}}^\Trans \in\mathbb{R}^{(\tau N_R)\times(\tau N_R)}. \label{eq:D_I}
\end{align}
Based on \eqref{eqn: prob_density_h} and \eqref{eqn: Prob_r_given_h_compact}, we rewrite \eqref{eqn: prob_r_first_appear} as
\begin{align}
\Pr(\mathbf{r}) &= \displaystyle \int_{\mathbb{C}^{N_TN_R}}\Pr(\mathbf{r}|\mathbf{h})\frac{e^{-\mathbf{h}^\Herm\boldsymbol{\Sigma}^{-1}\mathbf{h}}}{\pi^{N_TN_R}|\boldsymbol{\Sigma}|} {\rm d}\mathbf{h} \label{eqn: Prob_r} \\
& =\frac{1}{\pi^{\tau N_R}|\boldsymbol{\Omega}_{\rm b}|}
\int_{\mathbb{R}_+^{2\tau N_R}} e^{-\mathbf{z}^\Trans \mathbf{C}\mathbf{z}} {\rm d}\mathbf{z}, \label{eqn:Pr_r_compact}
\end{align}
with
\begin{align}
\mathbf{C} & = \begin{bmatrix}
\boldsymbol{\Lambda}_{\mathsf{R}}\mathbf{D}_{\mathsf{R}}\boldsymbol{\Lambda}_{\mathsf{R}} & \boldsymbol{\Lambda}_{\mathsf{R}}\mathbf{D}_{\mathsf{I}}^\Trans\boldsymbol{\Lambda}_{\mathsf{I}}\\
\boldsymbol{\Lambda}_{\mathsf{I}}\mathbf{D}_{\mathsf{I}}\boldsymbol{\Lambda}_{\mathsf{R}} & \boldsymbol{\Lambda}_{\mathsf{I}}\mathbf{D}_{\mathsf{R}}\boldsymbol{\Lambda}_{\mathsf{I}}
\end{bmatrix} \in\mathbb{R}^{(2\tau N_R)\times (2\tau N_R)}, \label{eqn: Matrix_C} \\
\mathbf{z} & = \left[\mathbf{x}^\Trans\;\;\mathbf{y}^\Trans\right]^\Trans \in\mathbb{R}^{2\tau N_R}. \label{eqn: vector_z}
\end{align}
Following similar steps, we rewrite the numerator of \eqref{eqn: h_MMSE_txt_formula} as
\begin{align}
\displaystyle\int_{\mathbb{C}^{N_T N_R}} \mathbf{h}\Pr(\mathbf{r}|\mathbf{h})p(\mathbf{h}){\rm d}\mathbf{h} & = \frac{\boldsymbol{\Sigma}\mathbf{A}^\Herm\boldsymbol{\Omega}_{\rm b}^{-1}\left[\boldsymbol{\Lambda}_{\mathsf{R}}\;\; j\boldsymbol{\Lambda}_{\mathsf{I}}\right]}{\pi^{\tau N_R}|\boldsymbol{\Omega}_{\rm b}|} \nonumber \\
& \phantom{=} \ \cdot \int_{\mathbb{R}_+^{2\tau N_R}}
\mathbf{z}e^{-\mathbf{z}^\Trans \mathbf{C}\mathbf{z}}{\rm d}\mathbf{z}.
\label{eqn: h_MMSE_numerator_final_compact}
\end{align}
Finally, based on \eqref{eqn:Pr_r_compact} and \eqref{eqn: h_MMSE_numerator_final_compact}, we express \eqref{eqn: h_MMSE_txt_formula} as
\begin{align}
\hspace{-2mm} \hat{\mathbf{h}}_{\rm MMSE} = \boldsymbol{\Sigma} \mathbf{A}^\Herm \boldsymbol{\Omega}_{\rm b}^{-1} \left[\boldsymbol{\Lambda}_{\mathsf{R}} \; \; j \boldsymbol{\Lambda}_{\mathsf{I}}\right] \left( \! \frac{\displaystyle \int_{\mathbb{R}_+^{2 \tau N_R}} \mathbf{z} e^{-\mathbf{z}^\Trans \mathbf{C} \mathbf{z}} {\rm d} \mathbf{z}}{\displaystyle \int_{\mathbb{R}_+^{2 \tau N_R}} e^{-{\mathbf{z}}^\Trans \mathbf{C} \mathbf{z}} {\rm d} {\mathbf{z}}} \! \right). \label{eqn: compact_MMSE_form1}
\end{align}
 
The fraction in \eqref{eqn: compact_MMSE_form1} denotes the mean of a truncated MVN distribution with $\mathbb{R}_{+}^{2\tau N_R}$ as its support. Moreover, since $\frac{{\rm d}}{{\rm d}\mathbf{z}} e^{-\mathbf{z}^\Trans \mathbf{C}\mathbf{z}}=-2\mathbf{Cz} e^{-\mathbf{z}^\Trans \mathbf{C}\mathbf{z}}$, we have
\begin{align}
\int_{\mathbb{R}_+^{2\tau N_R}} \mathbf{z} e^{-\mathbf{z}^\Trans \mathbf{C}\mathbf{z}} {\rm d}\mathbf{z} = \frac{1}{2}\mathbf{C}^{-1}\boldsymbol{\alpha},
\label{eqn: method_reduction_2}
\end{align}
where the $k$th element of $\boldsymbol{\alpha}\in\mathbb{R}^{2\tau N_R}$ is given by
\begin{align}
\alpha_k=\int_{\mathbb{R}_+^{2\tau N_R-1}} e^{-\mathbf{z}_{-k}^\Trans \mathbf{C}_{-k}\mathbf{z}_{-k}} {\rm d}\mathbf{z}_{-k}, \label{eqn: reduction_in_MMSE_numerator_by_integration}
\end{align}
with $\mathbf{z}_{-k}=[z_1\;\ldots \; z_{k-1}\; z_{k+1}\; \ldots \; z_{2\tau N_R}]^\Trans\in\mathbb{R}^{2\tau N_R-1}$ and where $\mathbf{C}_{-k}\in\mathbb{R}^{(2\tau N_R-1)\times (2\tau N_R-1)}$ is obtained from $\mathbf{C}$ by deleting its $k$th row and column. Hence, \eqref{eqn: compact_MMSE_form1} is equivalent to
\begin{align}
\hat{\mathbf{h}}_{\rm MMSE} = \frac{\boldsymbol{\Sigma}\mathbf{A}^\Herm\boldsymbol{\Omega}_{\rm b}^{-1} \left[\boldsymbol{\Lambda}_{\mathsf{R}}\;\; j\boldsymbol{\Lambda}_{\mathsf{I}}\right] \mathbf{C}^{-1}\boldsymbol{\alpha}}
{\displaystyle 2\int_{\mathbb{R}_+^{2\tau N_R}} e^{-\mathbf{z}^\Trans \mathbf{C}\mathbf{z}} {\rm d}\mathbf{z}}. \label{eqn: reduction_1}
\end{align}

\begin{remark}\label{remark: on pairwise_quantized_op_in_C}
Each off-diagonal element of $\mathbf{C}$ is affected by exactly two different elements from $[\mathbf{r}_{\mathsf{R}}^\Trans \;\;\mathbf{r}_{\mathsf{I}}^\Trans]
^\Trans$. The same comment also applies to $\mathbf{C}_{-k}$. This fact is instrumental for deriving computationally efficient expressions of $\hat{\mathbf{h}}_{\rm MMSE}$ under specific assumptions on the channel correlation or pilot symbols~\cite{Din24}.
\end{remark}

The expression in \eqref{eqn: reduction_1} is directly related to the orthant probability of a real-valued MVN distribution, defined as~\cite{Abrahamsom_64}
\begin{align}
\mathcal{P}(\boldsymbol{\Psi})= \frac{1}{(2\pi)^{\frac{L}{2}}\left|\boldsymbol{\Psi}\right|^{\frac{1}{2}}}\displaystyle \int_{\mathbb{R}_+^{L}} e^{-\frac{1}{2}\mathbf{u}^\Trans \boldsymbol{\Psi}^{-1}\mathbf{u}} {\rm d}\mathbf{u}, \label{eqn: orthant_Prob_formula}
\end{align}
where $\boldsymbol{\Psi} \in \Real^{L\times L}$ denotes the covariance matrix. Based on \eqref{eqn: orthant_Prob_formula}, we rewrite \eqref{eqn: reduction_in_MMSE_numerator_by_integration} as
\begin{align}
\alpha_k = \left(2\pi\right)^{\frac{2\tau N_R-1}{2}}\left|\frac{1}{2}\mathbf{C}_{-k}^{-1}\right|^{\frac{1}{2}}\mathcal{P}\left(\frac{1}{2}\mathbf{C}_{-k}^{-1}\right) \label{eqn: alpha_k_tmp}
\end{align}
and the integral in the denominator of \eqref{eqn: reduction_1} as
\begin{align}
\int_{\mathbb{R}_+^{2\tau N_R}} 
e^{-\mathbf{z}^\Trans \mathbf{C}\mathbf{z}} {\rm d}\mathbf{z}=\left(2\pi\right)^{\tau N_R}\left|\frac{1}{2}\mathbf{C}^{-1}\right|^{\frac{1}{2}}\mathcal{P}\left(\frac{1}{2}\mathbf{C}^{-1}\right).\label{eqn: h_MMSE_orthantP_tmp1}
\end{align}
It can also be shown that
\begin{align}
|\mathbf{C}_{-k}^{-1}|/|\mathbf{C}^{-1}|=|\mathbf{C}|/|\mathbf{C}_{-k}|=1/[\mathbf{C}^{-1}]_{kk}. \label{eqn: h_MMSE_orthantP_tmp2_det}
\end{align}
Based on \eqref{eqn: alpha_k_tmp}--\eqref{eqn: h_MMSE_orthantP_tmp2_det}, we rewrite \eqref{eqn: reduction_1} as 
\begin{align}
\hat{\mathbf{h}}_{\rm MMSE} = \frac{\boldsymbol{\Sigma}\mathbf{A}^\Herm\boldsymbol{\Omega}_{\rm b}^{-1} \left[\boldsymbol{\Lambda}_{\mathsf{R}} \; \; j \boldsymbol{\Lambda}_{\mathsf{I}}\right] \mathbf{C}^{-1}\left(\widetilde{\Diag{}}(\mathbf{C}^{-1})\right)^{-\frac{1}{2}}\mathbf{g}}
{2\sqrt{\pi}\mathcal{P}\left(\frac{1}{2}\mathbf{C}^{-1}\right)}, \label{eqn: reduction_2}
\end{align}
where the $k$th element of $\mathbf{g} \in \Real^{2 \tau N_R}$ is given by $g_k =\mathcal{P}\left(\frac{1}{2}\mathbf{C}_{-k}^{-1}\right)$.

\begin{remark} \label{rem:computation}
The channel covariance matrix $\boldsymbol{\Sigma}$ and the pilot matrix $\mathbf{S}$, which are embedded in $\boldsymbol{\Omega}_{\rm b}$ in \eqref{eqn: corr_mat_of_b} and in $\mathbf{C}$ in \eqref{eqn: Matrix_C} through \eqref{eq:D_R}--\eqref{eq:D_I}, determine the complexity of computing $\hat{\mathbf{h}}_{\rm MMSE}$ in \eqref{eqn: reduction_2}. The number of transmit antennas $N_T$ is not involved in the computation of the orthant probability.
\end{remark}


In~\cite{Childs-1967}, a method based on the characteristic function (CF), i.e., the Fourier transform with sign reversal, was used to express the orthant probability in \eqref{eqn: orthant_Prob_formula}. Based on~\cite{Abrahamsom_64, Childs-1967}, we have
\begin{align}
\mathcal{P}(\boldsymbol{\Psi}) = \begin{cases}
\frac{1}{4} + \frac{\arcsin(\psi_{12})}{2\pi} & \textrm{for } L=2, \\
\frac{1}{8} + \frac{\sum_{i=1}^2 \sum_{k=i+1}^3 \arcsin(\psi_{ik})}{4\pi} & \textrm{for } L=3,
\end{cases}
\label{eqn: OrthantP_N2_N3} 
\end{align}
where $\boldsymbol{\Psi}$ is standardized. In the Appendix, we extend the CF-based method to compute the numerator of the fraction in \eqref{eqn: compact_MMSE_form1}, thereby providing an alternative to \eqref{eqn: method_reduction_2}--\eqref{eqn: reduction_in_MMSE_numerator_by_integration}.

\section{Optimality of the BLMMSE Channel Estimator} \label{sec: cases_optimality}

In this section, building on the above framework and utilizing the observations in Remark~\ref{rem:computation}, we present three specific cases where the MMSE channel estimate is linear in $\mathbf{r}$, i.e., the BLMMSE channel estimator is optimal (Sections~\ref{sec_results: Uncorr_CH_Ortho_Pilot_mat}, \ref{sec_results:Corr_TX_EVD}, and~\ref{sec_results: Fully_Corr_N2}), along with one specific case where it is not (Section~\ref{sec: SIMO_tau_1}). Then, we determine a necessary and sufficient condition for the optimality of the BLMMSE channel estimator (Section~\ref{sec: optimality}). To facilitate the subsequent exposition, we first express the BLMMSE channel estimate as~\cite{Y_Li_et_al_BLMMSE}
\begin{align}
\hat{\mathbf{h}}_{\rm BLMMSE} & = \frac{\sqrt{\pi}}{2}\boldsymbol{\Sigma}\mathbf{A}^\Herm\mathbf{D}_\Omega^{-\frac{1}{2}}\Big(\arcsin\big(\mathbf{D}_\Omega^{-\frac{1}{2}}\Re\left(\boldsymbol{\Omega}_{\rm b}\right)\mathbf{D}_\Omega^{-\frac{1}{2}} \big) \nonumber \\
& \phantom{=} \ + j\arcsin\big(\mathbf{D}_\Omega^{-\frac{1}{2}}\Im\left(\boldsymbol{\Omega}_{\rm b}\right)\mathbf{D}_\Omega^{-\frac{1}{2}}
\big)\Big)^{-1}\mathbf{r},
\label{eqn: full_BLMMSE_formula}
\end{align}
with $\mathbf{D}_{\Omega} = \widetilde{\Diag{}}(\boldsymbol{\Omega}_{\rm b})$
and $\boldsymbol{\Omega}_{\rm b}$ defined in \eqref{eqn: corr_mat_of_b}.

\subsection{Linear Case: Uncorrelated Channel and Unitary Pilot Matrix} \label{sec_results: Uncorr_CH_Ortho_Pilot_mat}

Let $\tau=N_T$, and assume $\boldsymbol{\Sigma} = \mathbf{I}_{N_TN_R}$ and  $\mathbf{S} \mathbf{S}^\Herm = \eta \mathbf{I}_\tau$, with $\eta > 0$. Then,
\begin{align}
\mathbf{D}_{\mathsf{R}}=\frac{1}{\eta+\sigma^2}\mathbf{I}_{\tau N_R}, \quad \mathbf{D}_{\mathsf{I}} = \mathbf{0}, \quad \mathbf{C}=\frac{1}{\eta+\sigma^2}\mathbf{I}_{2\tau N_R}. \label{eqn: DR_DI_C_ortho_MIMO}
\end{align} Further algebraic manipulations based on \eqref{eqn: reduction_2} and \eqref{eqn: full_BLMMSE_formula} lead to 
\begin{align}
\hat{\mathbf{h}}_{\rm MMSE}= \frac{\left(\mathbf{S}^\Herm \otimes \mathbf{I}_{N_R}\right)\mathbf{r}}{\sqrt{\pi(\eta+\sigma^2)}}=\hat{\mathbf{h}}_{\rm BLMMSE}. \label{eqn: spatially_white_unitary_h_MMSE}
\end{align}

\vspace{-0.5mm}

This specific case is illustrated in Fig.~\ref{fig:fig_1} for different numbers of transmit antennas. The result in \eqref{eqn: spatially_white_unitary_h_MMSE} encompasses the uncorrelated SIMO system considered in~\cite{Utschick_etal_2023} as a special case and can be further generalized using the proposed framework, as shown next.

\subsection{Linear Case: Transmit-Only Channel Correlation} \label{sec_results:Corr_TX_EVD}

\begin{figure}
\centering
\begin{tikzpicture}

\begin{axis}[
	width=8cm,
	height=6.5cm,
	xmin=-10, xmax=30,
	ymin=0.35, ymax=0.6,
	xlabel={SNR [dB]},
	ylabel={MSE per antenna},
	ytick={0.35,0.4,0.45,0.5,0.55,0.6},
	xlabel near ticks,
	ylabel near ticks,
	x label style={font=\footnotesize},
	y label style={font=\footnotesize},
	ticklabel style={font=\footnotesize},
	legend style={at={(0.98,0.98)}, anchor=north east},
	legend style={font=\scriptsize, inner sep=1pt, fill opacity=0.75, draw opacity=1, text opacity=1},
	legend cell align=left,
	grid=both,
	title style={font=\scriptsize},
]

\addplot[thick, black]
table [x=SNR_dB, y=MSE_theo, col sep=comma] {figures/files_txt/MISO_TX_M8_0corr.txt};
\addlegendentry{$N_T = 8$ (uncorrelated)};

\addplot[thick, red]
table [x=SNR_dB, y=MSE_theo, col sep=comma] {figures/files_txt/MISO_TX_M16_0corr.txt};
\addlegendentry{$N_T = 16$ (uncorrelated)};

\addplot[thick, blue]
table [x=SNR_dB, y=MSE_theo, col sep=comma] {figures/files_txt/MISO_TX_M32_0corr.txt};
\addlegendentry{$N_T = 32$ (uncorrelated)};

\addplot[thick, black, dashed]
table [x=SNR_dB, y=MSE_theo, col sep=comma] {figures/files_txt/MISO_TX_M8_05_pi_6.txt};
\addlegendentry{$N_T = 8$ (correlated)};

\addplot[thick, red, dashed]
table [x=SNR_dB, y=MSE_theo, col sep=comma] {figures/files_txt/MISO_TX_M16_05_pi_6.txt};
\addlegendentry{$N_T = 16$ (correlated)};

\addplot[thick, blue, dashed]
table [x=SNR_dB, y=MSE_theo, col sep=comma] {figures/files_txt/MISO_TX_M32_05_pi_6.txt};
\addlegendentry{$N_T = 32$ (correlated)};

\addplot[thick, black, only marks, mark=o]
table [x=SNR_dB, y=MSE_opt, col sep=comma] {figures/files_txt/MISO_TX_M8_0corr.txt};

\addplot[thick, red, only marks, mark=o]
table [x=SNR_dB, y=MSE_opt, col sep=comma] {figures/files_txt/MISO_TX_M16_0corr.txt};

\addplot[thick, blue, only marks, mark=o]
table [x=SNR_dB, y=MSE_opt, col sep=comma] {figures/files_txt/MISO_TX_M32_0corr.txt};

\addplot[thick, black, only marks, mark=asterisk]
table [x=SNR_dB, y=MSE_opt, col sep=comma] {figures/files_txt/MISO_TX_M8_05_pi_6.txt};

\addplot[thick, red, only marks, mark=asterisk]
table [x=SNR_dB, y=MSE_opt, col sep=comma] {figures/files_txt/MISO_TX_M16_05_pi_6.txt};

\addplot[thick, blue, only marks, mark=asterisk]
table [x=SNR_dB, y=MSE_opt, col sep=comma] {figures/files_txt/MISO_TX_M32_05_pi_6.txt};

\end{axis}

\end{tikzpicture}
\caption{MSE per antenna versus SNR for the specific cases of uncorrelated channel and unitary pilot matrix described in Section~\ref{sec_results: Uncorr_CH_Ortho_Pilot_mat} (uncorrelated) and transmit-only channel correlation described in Section~\ref{sec_results:Corr_TX_EVD} (correlated). The lines and markers denote the analytical results and simulations, respectively.}
\label{fig:fig_1}
\end{figure}
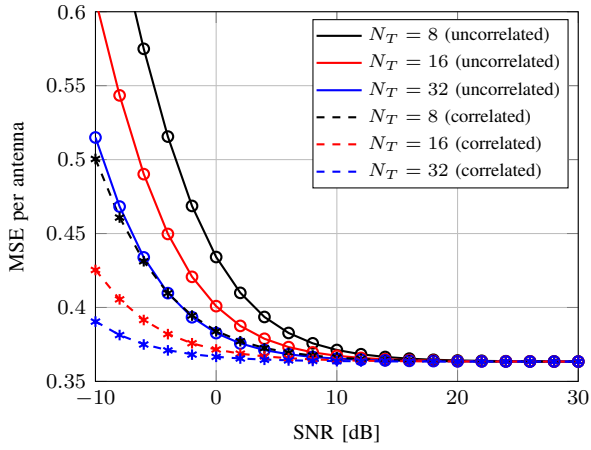

Let $\tau=N_T$ and assume $\mathbf{H}=\mathbf{H}_{\rm w}\boldsymbol{\Sigma}_{\rm TX}^{\frac{1}{2}}$, where $\boldsymbol{\Sigma}_{\rm TX} \in \Compl^{N_T \times N_T}$ represents the transmit channel correlation matrix and where the elements of $\mathbf{H}_{\rm w} \in \Compl^{N_R \times N_T}$ are i.i.d. $\mathcal{CN}(0, 1)$ random variables. Then, we have $\boldsymbol{\Sigma}=\boldsymbol{\Sigma}_{\rm TX}\otimes\mathbf{I}_{N_R}$. Now, let $\boldsymbol{\Sigma}_{\rm TX} = \mathbf{U} \boldsymbol{\Xi} \mathbf{U}^\Herm$ denote the eigenvalue decomposition of $\boldsymbol{\Sigma}_{\rm TX}$, with $\boldsymbol{\Xi} = \Diag{}(\left[\xi_i\right]_{i=1}^{N_T})$ and $\xi_i>0$. Utilizing the insights from Remark~\ref{rem:computation}, we set $\mathbf{S}=\sqrt{\eta}\mathbf{U}^\Herm$, which yields
\begin{align}
\mathbf{D}_{\mathsf{R}} = \Diag{\left(\left[\frac{1}{\eta\xi_i+\sigma^2}\right]_{i=1}^{N_T}\right)} \otimes \mathbf{I}_{N_R}, \quad \mathbf{D}_{\mathsf{I}} = \mathbf{0}.
\end{align}
Furthermore,
\begin{align}
\mathbf{C} = \BDiag(\mathbf{D}_{\mathsf{R}}, \mathbf{D}_{\mathsf{R}}) \label{eqn: C_TX_corr}
\end{align}
is also diagonal and the integrals in \eqref{eqn: compact_MMSE_form1} are over scalar variables, which allows us to write
\begin{align}
\hat{\mathbf{h}}_{\rm MMSE} & = \left(\left(\mathbf{U}\;\Diag{}\left(\left[\frac{\xi_i\sqrt{\eta}}{\sqrt{\eta\xi_i+\sigma^2}} \right]_{i=1}^{N_T}\right)\right)\otimes\mathbf{I}_{N_R}\right)\frac{\mathbf{r}}{\sqrt{\pi}} \\
& = \hat{\mathbf{h}}_{\rm BLMMSE}.\label{eqn: TX_corr_equivalence}
\end{align}

This specific case is included in Fig.~\ref{fig:fig_1} for different numbers of transmit antennas, where the transmit channel correlation coefficients are given by $[\boldsymbol{\Sigma}_{\rm TX}]_{ik} = \textsf{J}_0 (2 \pi (k-i) \Delta \gamma_{\rm max} \cos(\theta)) e^{-j 2 \pi (k-i) \Delta \sin(\theta)}$, where $\textsf{J}_0(\cdot)$ denotes the zeroth-order Bessel function of the first kind and where we have set $\gamma_{\rm max} = 0.1$ (maximum angle spread), $\theta = \frac{\pi}{6}$ (angle of arrival), and $\Delta = 0.5$ wavelengths (antenna spacing)~\cite{Abdi_2002}. Clearly, \eqref{eqn: spatially_white_unitary_h_MMSE} represents a special case of \eqref{eqn: TX_corr_equivalence} with $\boldsymbol{\Sigma}_{\rm TX}=\mathbf{I}_{N_T}$. If $\boldsymbol{\Sigma}_{\rm TX}$ is circulant, then a discrete Fourier transform matrix can be used as the pilot matrix.

\subsection{Linear Case: SIMO System with Real-Valued Channel Correlation ($N_R=2$)} \label{sec_results: Fully_Corr_N2}

Assume a correlated SIMO system with a single complex-valued pilot symbol $s$, i.e., $\tau=N_T=1$, so that the received signal in \eqref{sm: vec_SM} becomes
\begin{align}
\mathbf{b} = \mathbf{h} s + \mathbf{n} \in \Compl^{N_R}. \label{eqn: SIMO_1}
\end{align}
Let $\boldsymbol{\Sigma}$ be real-valued, as for example in the exponentially correlated channel model~\cite{Chiani_exp_model}. Then, we have
\begin{align}
\mathbf{D}_{\mathsf{R}} = \boldsymbol{\Omega}_{\rm b}^{-1}=(|s|^2\boldsymbol{\Sigma}+\sigma^2\mathbf{I}_{N_R})^{-1}, \quad \mathbf{D}_{\mathsf{I}}=\mathbf{0}, \label{eqn: SIMO_2}
\end{align}
and
\begin{align}\mathbf{C} = \BDiag(\boldsymbol{\Lambda}_{\mathsf{R}} \mathbf{D}_{\mathsf{R}} \boldsymbol{\Lambda}_{\mathsf{R}}, \boldsymbol{\Lambda}_{\mathsf{I}} \mathbf{D}_{\mathsf{R}} \boldsymbol{\Lambda}_{\mathsf{I}}).
\label{eqn: mat_C_SIMO}
\end{align}
When $N_R=2$, the diagonal blocks in \eqref{eqn: mat_C_SIMO} are $2 \times 2$ matrices. Finally, assuming a standardized $\boldsymbol{\Sigma}$ with channel correlation coefficient $[\boldsymbol{\Sigma}]_{12}=\rho_{12}$ and defining $\beta_{12}= \frac{\rho_{12}|s|^2}{|s|^2+\sigma^2}$, we obtain
\begin{align}
\hspace{-1mm} \hat{\mathbf{h}}_{\rm MMSE} = \frac{s^*\boldsymbol{\Sigma}\mathbf{T}^{-1}\mathbf{r}}{\sqrt{\pi(|s|^2+\sigma^2)}}
=\hat{\mathbf{h}}_{\rm BLMMSE}. \label{eqn: SIMO_N_eq_2}
\end{align}
with $[\mathbf{T}]_{12} = [\mathbf{T}]_{21}=\frac{2}{\pi}\arcsin(\beta_{12})$ and $[\mathbf{T}]_{11} = [\mathbf{T}]_{22}=1$.

This specific case is included in Fig.~\ref{fig:fig_2}, where the channel correlation coefficient between antennas $i$ and $k$ is given by $[\boldsymbol{\Sigma}]_{ik} = \rho^{|i-k|}$~\cite{Chiani_exp_model}.

\subsection{Non-Linear Case: SIMO System with Real-Valued Channel Correlation ($N_R=3$)} \label{sec: SIMO_tau_1}

\begin{figure}
\centering
\pgfdeclarelayer{background}
\pgfdeclarelayer{foreground}
\pgfsetlayers{background,main,foreground}

\begin{tikzpicture}

\begin{pgfonlayer}{background}
\begin{axis}[
	width=8cm,
	height=6.5cm,
	xmin=-10, xmax=30,
	ymin=0.25, ymax=0.9,
	xlabel={SNR [dB]},
	ylabel={MSE per antenna},
	ytick={0.1,0.2,0.3,0.4,0.5,0.6,0.7,0.8,0.9},
	xlabel near ticks,
	ylabel near ticks,
	x label style={font=\footnotesize},
	y label style={font=\footnotesize},
	ticklabel style={font=\footnotesize},
	legend style={at={(0.98,0.98)}, anchor=north east},
	legend style={font=\scriptsize, inner sep=1pt, fill opacity=0.75, draw opacity=1, text opacity=1},
	legend cell align=left,
	grid=both,
	title style={font=\scriptsize},
]

\addplot[thick, black]
table [x=SNR_dB, y=MSE_theo, col sep=comma] {figures/files_txt/N2_SIMO_035.txt};
\addlegendentry{$N_R = 2$ (MMSE = BLMMSE)};

\addplot[thick, black, dashed]
table [x=SNR_dB, y=MSE_opt, col sep=comma] {figures/files_txt/N3_SIMO_035.txt};
\addlegendentry{$N_R = 3$ (MMSE)};

\addplot[thick, black, only marks, mark=triangle]
table [x=SNR_dB, y=MSE_BLM, col sep=comma] {figures/files_txt/N3_SIMO_035.txt};
\addlegendentry{$N_R = 3$ (BLMMSE)};

\addplot[thick, red]
table [x=SNR_dB, y=MSE_theo, col sep=comma] {figures/files_txt/N2_SIMO_065.txt};

\addplot[thick, blue]
table [x=SNR_dB, y=MSE_theo, col sep=comma] {figures/files_txt/N2_SIMO_095.txt};

\addplot[thick, red, dashed]
table [x=SNR_dB, y=MSE_opt, col sep=comma] {figures/files_txt/N3_SIMO_065.txt};

\addplot[thick, red, only marks, mark=triangle]
table [x=SNR_dB, y=MSE_BLM, col sep=comma] {figures/files_txt/N3_SIMO_065.txt};

\addplot[thick, blue, dashed]
table [x=SNR_dB, y=MSE_opt, col sep=comma] {figures/files_txt/N3_SIMO_095.txt};

\addplot[thick, blue, only marks, mark=triangle]
table [x=SNR_dB, y=MSE_BLM, col sep=comma] {figures/files_txt/N3_SIMO_095.txt};

\draw (4,0.525) ellipse (0.5 and 0.02);
\draw (4,0.47) ellipse (0.5 and 0.025);
\draw (-4,0.665) ellipse (0.5 and 0.05);

\begin{scope}[>=latex]
\draw[-] (4,0.545) -- (5,0.625) {};
\draw[-] (3.5,0.47) -- (-5,0.375) {};
\draw[-] (-4,0.615) -- (-5,0.475) {};
\end{scope}

\node[black, font=\scriptsize] at (5,0.65) {$\rho = 0.35$};
\node[black, font=\scriptsize] at (-5,0.35) {$\rho = 0.65$};
\node[black, font=\scriptsize] at (-5,0.45) {$\rho = 0.95$};

\coordinate (zoom_coord) at (34,0.825);
\draw[black] (17.5,0.29) rectangle (22.5,0.37) {};
\begin{scope}[>=latex]
\draw[->] (axis cs:22.5,0.37) -- (axis cs:23.52,0.46) {};
\end{scope}

\end{axis}
\end{pgfonlayer}

\begin{pgfonlayer}{foreground}
\begin{axis}[
	axis background/.style={fill=white},
	at={(zoom_coord)},
	anchor={outer north east},
	width=0.18\textwidth,
	height=0.24\textwidth,
	xmin=17.5, xmax=22.5,
	ymin=0.29, ymax=0.37,
	ticks=none,
]

\addplot[thick, black]
table [x=SNR_dB, y=MSE_theo, col sep=comma] {figures/files_txt/N2_SIMO_035.txt};

\addplot[thick, black, dashed]
table [x=SNR_dB, y=MSE_opt, col sep=comma] {figures/files_txt/N3_SIMO_035.txt};

\addplot[thick, black, only marks, mark=triangle]
table [x=SNR_dB, y=MSE_BLM, col sep=comma] {figures/files_txt/N3_SIMO_035.txt};

\addplot[thick, red]
table [x=SNR_dB, y=MSE_theo, col sep=comma] {figures/files_txt/N2_SIMO_065.txt};

\addplot[thick, blue]
table [x=SNR_dB, y=MSE_theo, col sep=comma] {figures/files_txt/N2_SIMO_095.txt};

\addplot[thick, red, dashed]
table [x=SNR_dB, y=MSE_opt, col sep=comma] {figures/files_txt/N3_SIMO_065.txt};

\addplot[thick, red, only marks, mark=triangle]
table [x=SNR_dB, y=MSE_BLM, col sep=comma] {figures/files_txt/N3_SIMO_065.txt};

\addplot[thick, blue, dashed]
table [x=SNR_dB, y=MSE_opt, col sep=comma] {figures/files_txt/N3_SIMO_095.txt};

\addplot[thick, blue, only marks, mark=triangle]
table [x=SNR_dB, y=MSE_BLM, col sep=comma] {figures/files_txt/N3_SIMO_095.txt};

\end{axis}
\end{pgfonlayer}

\end{tikzpicture}
\caption{MSE per antenna versus SNR for the specific cases of SIMO with real-valued channel correlation and $N_R = 2$ described in Section~\ref{sec_results: Fully_Corr_N2} and $N_R = 3$ described in Section~\ref{sec: SIMO_tau_1}.}
\label{fig:fig_2}
\end{figure}
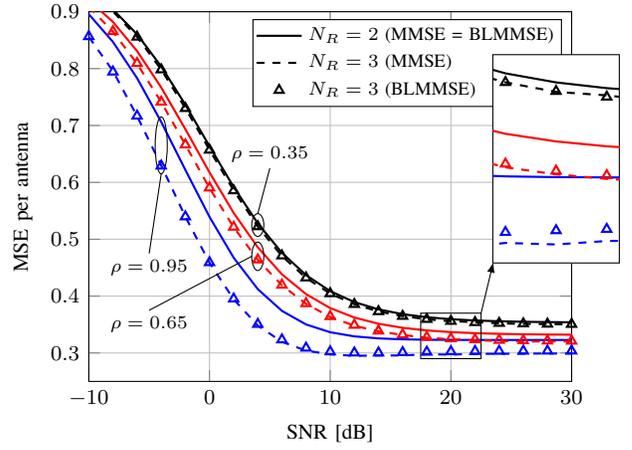

Here, we consider the same SIMO system described in Section~\ref{sec_results: Fully_Corr_N2} but with $N_R=3$. In this case, \eqref{eqn: SIMO_1}--\eqref{eqn: mat_C_SIMO} remain valid. Without loss of generality, assume a standardized $\boldsymbol{\Sigma}$ with channel correlation coefficient between antennas $i$ and $k$ denoted by $[\boldsymbol{\Sigma}]_{ik} = \rho_{ik}$. Further define $\beta_{ik}  = \frac{\rho_{ik}|s|^2}{|s|^2+\sigma^2}$. Based on \eqref{eqn: compact_MMSE_form1} or \eqref{eqn: reduction_2} and using \eqref{eqn: OrthantP_N2_N3}, we obtain the following result. The proof is omitted due to space limitations and can be obtained from~\cite{Din24}.

\begin{theorem} \label{theo: SIMO_N3_nonL}
Consider the SIMO system with $\tau = N_T = 1$ described by \eqref{eqn: SIMO_1}. Assume real-valued channel correlation and $N_R=3$. Then, we have
\begin{align}
\hat{\mathbf{h}}_{\rm MMSE}= \frac{s^*\boldsymbol{\Sigma}}{2\sqrt{\pi \left(|s|^2+\sigma^2\right)}}\left(\frac{\mathbf{v}_{\rm 3R}}{P_{\rm 3R}}+j\frac{\mathbf{v}_{\rm 3I}}{P_{\rm 3I}}\right), \label{eqn: h_mmse_N3}
\end{align}
with
\begin{align}
\hspace{-2mm} \mathbf{v}_{\rm 3R} & = \begin{bmatrix}
\frac{r_{\mathsf{R},1}}{4}+\frac{r_{\mathsf{R},1}r_{\mathsf{R},2}r_{\mathsf{R},3}}{2\pi}\arcsin\left(\frac{\beta_{23}-\beta_{12}\beta_{13}}{\sqrt{1-\beta_{12}^2}\sqrt{1-\beta_{13}^2}}\right) \\
\frac{r_{\mathsf{R},2}}{4}+\frac{r_{\mathsf{R},1}r_{\mathsf{R},2}r_{\mathsf{R},3}}{2\pi}\arcsin\left(\frac{\beta_{13}-\beta_{12}\beta_{23}}{\sqrt{1-\beta_{12}^2}\sqrt{1-\beta_{23}^2}}\right) \\ 
\frac{r_{\mathsf{R},3}}{4}+\frac{r_{\mathsf{R},1}r_{\mathsf{R},2}r_{\mathsf{R},3}}{2\pi}\arcsin\left(\frac{\beta_{12}-\beta_{13}\beta_{23}}{\sqrt{1-\beta_{13}^2}\sqrt{1-\beta_{23}^2}}\right)
\end{bmatrix}, \label{eqn: mean_vec_h_MMSE_SIMO_N3} \\
\hspace{-2mm} P_{\rm 3R} & = \frac{1}{8}+\frac{\sum_{i=1}^2\sum_{k=i+1}^3r_{\mathsf{R},i}r_{\mathsf{R},k}\arcsin( \beta_{ik})}{4\pi},
\label{eqn: ortP_R_h_MMSE_SIMO_N3}
\end{align}
and where $\mathbf{v}_{\rm 3I}$ (resp. $P_{\rm 3I}$) is the same as $\mathbf{v}_{\rm 3R}$ (resp. $P_{\rm 3R}$) except that $r_{\mathsf{R}, i}$ in \eqref{eqn: mean_vec_h_MMSE_SIMO_N3} (resp. \eqref{eqn: ortP_R_h_MMSE_SIMO_N3}) is replaced by $r_{\mathsf{I}, i}$.
\end{theorem}

\noindent Under the assumptions of Theorem~\ref{theo: SIMO_N3_nonL}, from \eqref{eqn: full_BLMMSE_formula}, we have
\begin{align}
\hat{\mathbf{h}}_{\rm BLMMSE}= \frac{s^*\boldsymbol{\Sigma}\widetilde{\mathbf{T}}^{-1}\mathbf{r}}{\sqrt{\pi \left(|s|^2+\sigma^2\right)}},
\label{eqn: h_blmmse_N3}
\end{align}
with $[\widetilde{\mathbf{T}}]_{ik} = \frac{2}{\pi}\arcsin(\beta_{ik})$ for $i\neq k$ and $[\widetilde{\mathbf{T}}]_{ii} = 1$.

\begin{remark} \label{rem: special_cases_N3} Note that \eqref{eqn: h_mmse_N3} is non-linear in $\mathbf{r}$ and thus not equivalent to \eqref{eqn: h_blmmse_N3}, unless one of the following holds: $\rho_{23}=\rho_{13}=0$, $\rho_{13}=\rho_{12}=0$, or $\rho_{12}=\rho_{23}=0$. 
\end{remark}

\begin{remark}
Theorem~\ref{theo: SIMO_N3_nonL} provides the simplest closed-form expression for the MMSE channel estimate when it is non-linear in $\mathbf{r}$.
\end{remark}

This specific case is included in Fig.~\ref{fig:fig_2}, which shows that the BLMMSE channel estimator is nearly optimal in cases with weak channel correlation.

\subsection{Necessary and Sufficient Optimality Condition} \label{sec: optimality}

Finally, we provide a necessary and sufficient condition for the optimality of the BLMMSE channel estimator. The proof builds on the intuition obtained from Sections~\ref{sec: framework}, \ref{sec_results: Fully_Corr_N2}, and~\ref{sec: SIMO_tau_1}; the details are omitted due to space limitations and can be found in~\cite{Din24}.

\begin{theorem} \label{theo: Theorem_BLMMSE_opt}
The BLMMSE channel estimator is optimal if and only if the off-diagonal elements of $\mathbf{C}$ in \eqref{eqn: Matrix_C} satisfy $[\mathbf{C}]_{il} \neq 0$ for at most one value of $l$ other than $l=i$, i.e., an element $z_i$ of the random vector $\mathbf{z}$ defined in \eqref{eqn: vector_z} is correlated with at most one other element $z_l$, for $l \neq i$.
\end{theorem}

\noindent It is straightforward to observe that $\mathbf{C}$ in \eqref{eqn: DR_DI_C_ortho_MIMO}, \eqref{eqn: C_TX_corr}, and \eqref{eqn: mat_C_SIMO} with $N_R=2$ satisfies the condition of Theorem~\ref{theo: Theorem_BLMMSE_opt} and, therefore, the BLMMSE channel estimator is optimal. On the other hand, \eqref{eqn: mat_C_SIMO} with $N_R=3$ does not satisfy the condition of Theorem~\ref{theo: Theorem_BLMMSE_opt} except for the special cases mentioned in Remark~\ref{rem: special_cases_N3}.

\begin{appendix}

Motivated by~\cite{Childs-1967}, we provide an alternative to \eqref{eqn: method_reduction_2}--\eqref{eqn: reduction_in_MMSE_numerator_by_integration} for computing the numerator of the fraction in \eqref{eqn: compact_MMSE_form1}. To this end, let $p(\mathbf{u})$ correspond to $\mathcal{N}(\mathbf{0}, \boldsymbol{\Psi})$ and write
\begin{align}
\displaystyle \int_{\mathbb{R}_+^{L}}\mathbf{u}p(\mathbf{u}){\rm d}\mathbf{u}=\displaystyle \int_{\mathbb{R}_+^{L}} \frac{\mathbf{u} e^{-\frac{1}{2}\mathbf{u}^\tran \boldsymbol{\Psi}^{-1}\mathbf{u}}}{(2\pi)^{\frac{L}{2}}\left|\boldsymbol{\Psi}\right|^{\frac{1}{2}}}  {\rm d}\mathbf{u}, \label{eqn: app_0evaluate_mean_truncated_Gaussian} 
\end{align}
whose $i$th element is given by 
\begin{align}
\mathcal{I}_i =\displaystyle \int_{\mathbb{R}_+^{L}} u_ip(\mathbf{u}) {\rm d}\mathbf{u} = \displaystyle \int_{\mathbb{R}^{L}} \underbrace{u_i\mathcal{U}(u_i)\prod_{\substack{k=1\\k\neq i}}^L \mathcal{U}(u_k)}_{\varphi_i(u_1, \ldots, u_L)=\varphi_i(\mathbf{u})} p(\mathbf{u}) {\rm d}\mathbf{u}, \label{eqn: app_2evaluate_mean_truncated_Gaussian} 
\end{align}
where $\mathcal{U}(\cdot)$ represents the Heaviside step function. Let $\boldsymbol{\omega}=[\omega_1\ldots\omega_L]^\tran$ and express the Fourier transform of $\varphi_i(\mathbf{u})$ as
\begin{align}
\mathcal{F}_{\varphi_i}(\boldsymbol{\omega})= \left(j\pi\delta'(\omega_i)-\frac{1}{\omega_i^2}\right)\prod_{\substack{k=1\\k\neq i}}^L\left(\pi\delta(\omega_k)+\frac{1}{j\omega_k}\right), \label{eqn: FT of var_step_fcns}
\end{align}
where $\delta(\cdot)$ is the Dirac delta function and $\delta'(\cdot)$ its derivative. Meanwhile, the Fourier transform of $p(\mathbf{u})$ is given by $e^{-\frac{1}{2}\boldsymbol{\omega}^\tran\boldsymbol{\Psi}\boldsymbol{\omega}}$.
As in~\cite{Childs-1967}, we use the Parseval–Plancherel identity to rewrite \eqref{eqn: app_2evaluate_mean_truncated_Gaussian} as
\begin{align}
\mathcal{I}_i=\frac{1}{(2\pi)^L}\displaystyle \int_{\mathbb{R}^{L}}  \mathcal{F}_{\varphi_i}(\boldsymbol{\omega})e^{-\frac{1}{2}\boldsymbol{\omega}^\tran\boldsymbol{\Psi}\boldsymbol{\omega}} {\rm d}\boldsymbol{\omega}. \label{eqn: app_3evaluate_mean_truncated_Gaussian}
\end{align} 
Combining \eqref{eqn: app_0evaluate_mean_truncated_Gaussian}, \eqref{eqn: FT of var_step_fcns}--\eqref{eqn: app_3evaluate_mean_truncated_Gaussian}, and~\cite[Eq. (6)]{Childs-1967} provides a method based solely on the CF to compute the fraction in \eqref{eqn: compact_MMSE_form1}.

As an example, consider the case of $L=2$ and assume a standardized $\boldsymbol{\Psi}$ with off-diagonal element $\psi_{12}$. 
Using \eqref{eqn: app_3evaluate_mean_truncated_Gaussian}, it can be shown that $\mathcal{I}_1=\mathcal{I}_2= (1+\psi_{12})/(2\sqrt{2\pi})$. The case of $L=3$ can be readily developed following similar steps.
\end{appendix}

\addcontentsline{toc}{chapter}{References}
\bibliographystyle{IEEEtran}
\bibliography{refs_abbr,refs}

\end{document}